\documentclass[12pt,draft]{article}

\usepackage{amsthm,amssymb,amsmath,amscd,amstext,comment}
\usepackage{latexsym,mathrsfs,mathptmx,xypic}

\newtheorem{definition}{Definition}[section]
\newtheorem{theorem}{Theorem}[section]
\newtheorem{lemma}{Lemma}[section]
\newtheorem*{remark}{{\it Remark}}

\newcommand{\nc}{\newcommand} 

\nc{\N}{{\mathbb N}}
\nc{\Z}{{\mathbb Z}}
\nc{\T}{{\mathbb T}}
\nc{\R}{{\mathbb R}}
\nc{\C}{{\mathbb C}}
\nc{\HH}{{\mathbb H}}

\nc{\dd}{{\rm d}}
\nc{\ii}{{\bf i}}

\nc{\ca}{{\mathscr A}}
\nc{\cg}{{\mathscr G}}
\nc{\cm}{{\mathscr N}}
\nc{\cs}{{\mathscr S}}

\begin{document}

\title{A set-theoretic analysis of the black hole entropy puzzle}

\author{G\'abor Etesi\\
\small{{\it Department of Algebra and Geometry, Institute of Mathematics,}}\\
\small{{\it Budapest University of Technology and Economics,}}\\
\small{{\it M\H uegyetem rkp. 3., H-1111 Budapest, Hungary}}
\footnote{E-mail: {\tt etesi@math.bme.hu, etesigabor@gmail.com}}}

\maketitle

\pagestyle{myheadings}
\markright{G. Etesi: A set-theoretic analysis of the black hole 
entropy puzzle}

\thispagestyle{empty}

\begin{abstract} Motivated by the known mathematical and physical problems 
arising from the current mathematical formalization of the physical 
spatio-temporal continuum, as a substantial technical clarification of our 
earlier attempt \cite{ete2}, the aim in this paper is twofold. Firstly, by 
interpreting Chaitin's variant of G\"odel's first incompleteness theorem 
as an inherent uncertainty or fuzziness present in the set of real numbers, a 
set-theoretic entropy is assigned to it using the Kullback--Leibler 
relative entropy of a pair of Riemannian manifolds. Then 
exploiting the non-negativity of this relative entropy an 
abstract Hawking-like area theorem is derived. 

Secondly, by analyzing Noether's theorem on 
symmetries and conserved quantities, we argue that whenever the four 
dimensional space-time continuum containing a black hole is modeled by the 
set of real numbers in the mathematical formulation of general relativity, 
the hidden set-theoretic entropy of this latter structure reveals itself as the 
entropy of the black hole (proportional to the area of its 
``instantaneous'' event horizon), indicating that this apparently 
physical quantity might have a pure set-theoretic origin, too. 
\end{abstract}

\centerline{PACS numbers: 01.55.+b; 02.10.-v; 03.75.Hh; 04.70.-s; 05.70.-a}
\centerline{Keywords: {\it Continuum; Chaitin incompleteness; 
Kullback--Leibler divergence; Black hole entropy}}


\section{Introduction}
\label{one}


The mathematical formalization of the continuum in its current form, 
what we shall call the {\it arithmetical continuum} or equivalently the 
{\it set of real numbers} $\R$ here, is widely used everywhere in mathematics, 
physics, engineering sciences, mathematical biology, economy and 
sociology, etc., etc. However, in spite of its success, the utilization 
of the arithmetical continuum leads to various difficulties both in pure 
mathematics \cite{wey} and its applications (for an excellent survey of 
physics see \cite{bae}); all of these difficulties basically originate 
from the fact that the arithmetical continuum has an inherent infinite 
transcendental structure in many aspects as a result of modeling the 
continuum i.e. something possesing {\it extension} by an unknownly huge 
cardinalty collection of extensionless {\it points}. Perhaps the most relevant 
question related with the problematics of the arithmetical continuum on the 
pure mathematical side is the (in our opinion) unsettled status of 
{\it Cantor}'s continuum hypothesis; while a simple but painful example from 
theoretical physics is the divergence of the total electric energy of an 
electrically charged {\it point} particle in Faraday--Maxwell 
electrodynamics (leading eventually to the complicated {\it 
renormalization} issues in classical and quantum field theories). 
Fortunately, by performing simple {\it experiments}, we know that this 
and the various other occurences of divergences in classical and quantum 
electrodynamics are nothing but artifacts originating from the 
mathematical formulation (which uses the arithmetical continuum) of 
these theories hence we can identify and isolate these divergences quite 
easily. But what about the singularities or other phenomena 
(mathematically) predicted by general relativity for instance? Lacking 
unambigous experimental evidences we cannot make a commitment about 
their ontological status yet.

The easiest way to get rid of the various divergence problems in physical 
theories arising from modeling mathematically the physical spatio-temporal 
continuum with the arithmetical one, is to declare that the mathematical 
description of space-time should be simply finite; that is the mathematical 
structure modeling physical space-time should have non-zero finite 
cardinality $0<N<+\infty$ as a set. Finiteness necessarily implies the 
geometric relation $N\sim V/V_{\rm Planck}$ where $V$ is the volume of a 
spatio-temporal region. In this framework it is yet reasonable to suppose that 
at macroscopic scales (i.e. as $V\rightarrow+\infty$) the cardinality of 
space-time is extremely huge, practically infinite hence the classical 
geometric description (provided by general relativity theory with its precise 
mathematics) applies as an approximation; nevertheless the macroscopic 
cardinality $N$ of the observable Universe is determined by some 
microscopic cardinality $0<N_{\rm Planck}<+\infty$ hence is finite. However 
as one approaches microscopic scales (i.e. as $V\rightarrow V_{\rm Planck}$) 
the finiteness gets more and more relevant and the supposed finite 
microscopic cardinality attained at the Planck scales. If one indeed 
wants to describe not merely a geometric continuum but the physical space-time 
itself then one also has to take into account that at microscopic scales 
the spatio-temporal continuum more-and-more resembles the vacuum state of a 
relativistic quantum field with its known microscopic properties  
(described by some yet mathematically problematic relativistic quantum field 
theory). Thus talking about the finiteness of space-time in fact 
means that one supposes that the physical vacuum has finite physical 
degrees of freedom. Consequently the cardinality $N$ of a volume $V$ containing 
vacuum is proportional to its energy content, i.e. expected to satisfy 
$N\sim E/E_{\rm Planck}$ too, where $E$ is the vacuum energy within $V$. 

The quantum vacuum is subject to {\it Heisenberg}'s various 
{\it uncertainty principles}. But if these are also involved in the 
description, the cardinality of the set modeling the physical 
space-time, if finite, gets problematic at the microscopic (hence 
the macroscopic) level. This is simply because as the size of a spatial volume 
approaches the Planck length or its time of existence gets very short, 
the fluctuation of the spatio-temporal 
cardinality gets comparable with the cardinality itself. More precisely we 
assumed that $N\sim V/V_{\rm Planck}\sim (L/\ell_{\rm Planck})^3$. Then 
using $\Delta L\Delta p\geqq\hbar$ the relative fluctuation of the 
vacuum cardinality within a volume $V$ is estimated from below as
\[\frac{\Delta N}{N}\sim\left(\frac{\Delta L}{L}\right)^3 
\geqq\left(\frac{\hbar}{L\Delta p}\right)^3\:\:.\]
On substituting for $L$ the fundamental length $\ell_{\rm Planck}$ and for 
$\Delta p$ the fundamental momentum uncertainty $p_{\rm Planck}= 
m_{\rm Planck}c$ we find 
\[\frac{\Delta N_{\rm Planck}}{N_{\rm Planck}}\geqq 
\left(\frac{\hbar}{\sqrt{\hbar G/c^3} \sqrt{\hbar c/G}\:c}\right)^3=1\]
in a small volume comparable to the Planck length. Likewise, based on the 
finiteness assumption, we know that $N\sim E/E_{\rm Planck}$ too. Putting this 
together with $\Delta E\Delta t\geqq\hbar$ gives 
\[\frac{\Delta N}{N}\sim\frac{\Delta E}{E}\geqq \frac{\hbar}{E\Delta t}\:\:.\] 
The fundamental energy of the vacuum is $E_{\rm Planck}=m_{\rm Planck}c^2$ and 
the minimal observation time of it is $t_{\rm Planck}$. Hence we obtain again 
\[\frac{\Delta N_{\rm Planck}}{N_{\rm Planck}}\geqq
\frac{\hbar}{c^2\sqrt{\hbar c/G} \sqrt{\hbar G/c^5}}=1\] 
shortly after the Big Bang for instance. Since the relative fluctuation is a 
dimensionless quantity, having unit magnitude means that it is meaningless to 
talk about a sharp (i.e. finite) cardinality $N_{\rm Planck}$ of the vacuum at 
short space or time scales, if quantum mechanics is also taken into account. 
This overall uncertainty might be the core physical reason why physical 
space-time is modeled by infinite mathematical structures: if one 
indeed wants to describe the physical space and time as the vacuum of a 
relativistic quantum field (and not merely as a geometric continuum) then 
using infinite sets is an apparently unavoidable mathematical way to grasp 
the overall spatio-temporal uncertainity of the relativistic quantum field 
comprising the true physical vacuum.

Having seen that modeling physical space-time with infinite mathematical 
structures is not easy to exclude, we return to our former question 
concerning general relativity: the purpose of this paper is to examine 
whether the several conceptional, technical, common sense, etc. 
controversies connected with {\it black hole entropy} 
\cite{bar-car-haw,haw1,haw2,str-vaf,wal1,wal3} do at least in part emanate 
from the fact that general relativity mathematically rests on 
the arithmetical continuum, an infinite structure? 

The paper is organized as follows. In Section \ref{two} we recall {\it 
Chaitin}'s reformulation of {\it G\"odel}'s first incompleteness theorem 
(see Theorem \ref{chaitin} here) and interpret its content---with hindsight 
dictated by {\it Heisenberg}' uncertainity---as the presence of 
an inherent uncertainty or fuzziness within the arithmetical continuum 
(which might also be a consequence that its final constituents are 
extensionless). This strongly motivates to introduce a 
statistico-physical analogy and to talk in this context about the pure 
``set-theoretic entropy'' of the arithmetical conntinuum. Quite 
interestingly this idea can be rigorously grasped by the aid of 
a key concept of current information theory, namely the 
{\it Kullback--Leibler relative entropy} or {\it divergence} 
\cite[Chapter 8]{cov-tho} adapted to a pair of Riemannian manifolds 
(see Theorem \ref{kl} here). Computing this quantity over a compact 
manifold-with-boundary and 
exploiting its non-negativity an abstract Riemannian geometric analogue 
of {\it Hawking}'s area theorem \cite{haw1} is obtained (Theorem 
\ref{hawking} here). Then, in Section \ref{three} basically following 
\cite[Section 3]{ete2} we recall and refine {\it Noether}'s theorem on 
symmetries and conserved quantities and prove that within general 
relativity a conserved quantity to diffeomorphisms can be assigned which 
is not zero if a stationary black hole is present and can be identified 
with its entropy (proportional to the area of the ``instantaneous'' 
event horizon \cite{haw1}). Then we argue, based on their common 
diffeomorphism invariance, that the ``set-theoretic'' and the black hole 
entropies are strongly related. This suggests that the latter entropy 
might be a consequence of the former one as a result of modeling general 
relativity mathematically over the arithmetical continuum therefore the 
long-sought physical degrees of freedom responsible for black hole 
entropy have at least in part simply a pure mathematical origin only, 
corresponding to the artificial division of the space-time continuum 
into points (see the discussion at the end of Section \ref{three}). 

Finally Section \ref{four} is an Appendix in which we make an attempt 
(certainly incomplete at this stage of the art) to introduce a temporal 
structure in a covariant way into general relativity: recall that 
although being our most advanced theory dealing with the structure 
of space and time, general relativity is in fact a ``timeless theory'' 
due to its diffeomorphism invariance (cf. e.g. \cite[Chapter 3]{cal}). 


\section{Chaitin incompleteness and the entropy of the continuum}
\label{two}


Accepting the structure of the arithmetical continuum or equivalently 
the set $\R$ of real numbers,\footnote{Here by a {\it real number} $x$ 
we mean by definition a Dedekind slice or cut of the rationals and by 
the {\it set of real numbers} $\R$ the collection of all of them. $\R$ 
can be equipped with the usual structures (addition, multiplication and 
ordering) and posssesses the completeness property rendering it a 
complete ordered field. It is well-known that $\R$ as a complete ordered 
field (formulable in a second order theory only!) is essentially 
unique.} it is the totality of, or more precisely the disjoint union of, 
its individual constituents called {\it real numbers} or---speaking 
geometrically---its {\it points}: 
\[\R =\bigsqcup_{x\in\R}\{x\}\:\:\:.\] 
If indeed this is the optimal mathematical structure of the continuum 
(e.g. it is possible to construct it without points \cite{hel-sha}), 
then one would expect that using mathematical tools only, one is able to 
``locate'' the individual constituents $x$ within their totality $\R$ or 
equivalently, to make a mathematical distinction between them in an 
effective way. We know from our elementary university studies that upon 
fixing a notational convention every real number admits a well-defined 
(for instance decimal) expansion which means that this expansion exists 
for all real numbers and is unique in the sense that two expansions 
coincide if and only if the corresponding two real numbers are equal. It 
is however a quite surprising observation that in general the existence 
of this well-defined and unique expansion is the only available property 
of a ``truely generic'' real number i.e. a typical element $x\in\R$. 
Therefore our question about an effective mathematical way of 
``picking'' a single element $x\in\R$ is in fact a question about the 
effectiveness of making distinctions between generic real numbers in 
terms of their (decimal) expansions.

To approach this problem first let us recall the idea of Kolmogorov 
complexity or algorithmic compressibility or computability of a real 
number (cf. e.g. \cite[Chapter 14]{cov-tho}). Let $x\in\R$ be given and denote 
by ${\bf T}_x$ the (probably empty) set of those Turing machines which 
reproduce $x$ in the following way: if ${\bf T}_x\not=\emptyset$ and some 
$T_x\in {\bf T}_x$ is given with an input $n\in\N$ then the output 
$T_x(n)\in\N$ consists of e.g. precisely the first $n$ digits of the 
expansion of $x$. Denoting by $\vert T_x\vert\in\N$ the length of the Turing 
machine (considered as an algorithm or program in some programming language) 
define 
\[K(x):=\left\{\begin{array}{ll}
                     +\infty & \mbox{if\:\:\:${\bf T}_x=\emptyset$}\\
                             & \\
                     \inf\limits_{T_x\in{\bf T}_x}
                                            \vert T_x\vert <+\infty &
                                       \mbox{if\:\:\:${\bf T}_x\not=\emptyset$}
               \end{array}\right.\] 
and call the resulting (extended) natural number 
$K(x)\in\N\cup\{+\infty\}$ the {\it Kolmogorov complexity} of $x\in\R$. 
Then $K(x)=+\infty$ corresponds to the situation when no algorithms 
reproducing $x$ in the above sense exist hence $x$ is not 
{\it algorithmically compressible} or not {\it computable} (by simple 
cardinality arguments the vast majority of real numbers belongs to this 
class including, as a yet definable hence mild example, Chaitin's 
famous $\Omega$ number \cite[Section 14.8]{cov-tho}) while $K(x)<+\infty$ 
corresponds to the opposite case (containing all familiar 
real numbers like $7$, $\frac{5}{8}$, $\sqrt{2}$, $\pi$, ${\rm e}$,...). 
It is clear that the only important question about $x$ in this context is 
whether $K(x)=+\infty$ or $K(x)<+\infty$ and in the latter case only the 
magnitude of $K(x)$ is relevant, for its particular value depends on the 
details of the sort of expansion of $x$, the programming language for $T_x$, 
etc. hence does not carry essential information. A remarkable observation 
about Kolmogorov complexity is the following result:

\begin{theorem}{\rm (Chaitin's version of G\"odel's first 
incompleteness theorem \cite{cha})} For any (sufficiently rich, consistent, 
recursively enumerable) axiomatic system $S$ 
based on a first order language $L$ there exists a natural number 
$0<N_S<+\infty$ such that there exists no real number $x$ for which the 
proposition 
\[K(x)\geqq N_S\]
is provable within $S$.\hspace{5.3in}$\square$
\label{chaitin}
\end{theorem}

\noindent Motivated in various ways by \cite{ger-har, sch1, sch2} we 
interpret this quite surprising mathematical fact from our 
viewpoint as follows: taking into account that the only known property of 
a generic real number which fully characterizes it is its existing (decimal) 
expansion, but the Kolmogorov complexity of this expansion hence the 
expansion itself generally is not fully determinable (by proving 
theorems on it in an axiomatic system), 
there is in general no way, using standard mathematical tools in the 
broadest sense, to ``sharply pick'' any element from the arithmetical 
continuum. Consequently, from the viewpoint of an ``effective mathematical 
activity'', {\it the structure of the arithmetical continuum i.e. the set 
$\R$ of real numbers contains an inherent uncertainty or fuzziness in the 
sense that its individual disjoint constituents cannot be distinguished 
from each other in a universal and effective mathematical way}. 

The above interpretation of Theorem \ref{chaitin} serves as a motivation 
to introduce a {\it statistico-physical analogy} for the arithmetical 
continuum offering a fresh look into its structure. 
First, generalizing the above decomposition of the real line into its 
points, we accept as usual that every (finite dimensional, real) differentiable 
manifold $M$ admits a decomposition into its disjoint constituent points: 
\begin{equation}
M=\bigsqcup_{x\in M}\{x\}\:\:\:.
\label{szetszedes}
\end{equation}
Speaking intutively, we can make three observations about this 
decomposition: all the points $x$ of $M$ (i) are homologous i.e. ``look the 
same'', (ii) are terminal objects i.e. they do not possess 
any further internal structure nevertheless their collection 
gives back precisely $M$ and (iii) are disjoint from each other i.e. they 
``do not interact''. Except its cardinality this decomposition of $M$ 
therefore strongly resembles the structure of an ideal gas as usually 
defined in statistical physics. Take an 
{\it abstract set} $X$ whose cardinality coincides with that of the continuum 
in ZFC set theory and regard it as an {\it abstract ideal gas} $X$ such that 
its {\it elements} correspond to the {\it atoms} of the ideal gas $X$. 
Extending this analogy further, the left hand side of (\ref{szetszedes}) i.e. 
a differentiable manifold $M$ with its global topological, smooth, 
etc. ``macroscopic'' properties can be regarded as one possible 
{\it macrostate of $X$} while the right hand side of 
(\ref{szetszedes}) i.e. the particular identification of $M$ with its elements 
as a particular {\it microstate of $X$} within the macrostate $M$. The 
{\it equilibrium dynamics of $X$} in its macrostate $M$ is generated by 
{\it diffeomorphisms}; hence another microstate of $X$ within the same 
macrostate $M$ is achieved by picking any diffeomorphism $f: M\rightarrow M$ 
and writing $f(M)=M$ again, and then taking the corresponding new 
decomposition  
\begin{equation}
M=\bigsqcup\limits_{x\in M}\{f(x)\}\:\:.
\label{fszetszedes}
\end{equation}
Another differentiable manifold $N$ not diffeomorphic to $M$ (in the broadest 
sense i.e. possibly having different dimension, number of connected 
components, etc.) might be interpreted as a different macrostate (with 
its corresponding assembly of microstates created by diffeomorphisms) of 
the same astractly given ideal gas $X$. However this abstract ideal gas can 
even appear in completely different i.e. non-geometric, discontinuous 
macrostates as well like e.g. in the form of the 
{\it Cantor set} $C\subset\R$ or some other abstract {\it topological 
space} (with its {\it homeomorphisms} creating the corresponding assembly of 
microstates), or just simply in the form of some {\it set} 
(together with its {\it bijections}), etc., etc. 

In accord with this analogy Theorem \ref{chaitin} is interpreted 
as a fundamental result about the indistinguishability of the individual 
microstates of $X$ realizing the same macrostate $M$. The next 
standard step in statistical mechanics is to introduce a tool, a 
measure, capable to capture the amount of information loss created by the 
passage from individual microstates to their common macrostate. This 
measure is known as the {\it entropy} of the ideal gas in a given macrostate. 
How could we characterize this entropy within our analogy? Proceeding 
completely formally along the way of Boltzmann's classical approach to 
entropy we can argue as follows. Certainly all possible microstates of the 
abstract ideal gas $X$ are parameterized by the elements of the group of its 
all set-theoretic bijections ${\rm Bij}(X)$ while its possible 
microstates within the macrostate $M$ are parameterized by its subgroup of 
diffeomorphisms ${\rm Diff}(M)$. Therefore restricting the dynamics of $X$ 
from ${\rm Bij}(X)$ to ${\rm Diff}(M)$ {\it by construction} means that $M$ 
is an equilibrium state of $X$. If we further assume 
that all microstates appear with equal probability i.e. a sort of ergodicity 
holds for $X$ then as a first trial we {\it formally} put 
\[\mbox{Entropy of the set $X$ in its manifold-macrostate $M$}\:\sim\:
\log\Gamma({\rm Diff}(M))\]
with $\Gamma$ being an, at this state of the art admittedly 
hypothetical, volume measure on ${\rm Bij}(X)$ 
depending on a particular choice of the 
axiomatic system $S$ in Theorem \ref{chaitin}. Taking into account 
that if $M$ has positive dimension then 
$1\subsetneqq {\rm Diff}(M)\subsetneqq {\rm Bij}(X)$, we expect 
$1\lvertneqq\Gamma({\rm Diff}(M))\lvertneqq\Gamma({\rm Bij}(X))$ to 
hold such that the resulting entropy expression is a finite positive number 
and at least in its magnitude being independent of any choice for $S$ as 
dictated by the universality of Theorem \ref{chaitin}. Note that despite 
being formally ill-defined, {\it by construction} this entropy formula is 
invariant under diffeomorphisms of $M$ since a diffeomorphism does not 
change the given macrostate. 

Let us try to grasp this set-theoretic entropy more 
precisely from a mathematical viewpoint. To achieve this we follow 
\cite[Chapter 8]{cov-tho} and introduce the Kullback--Leibler 
relative entropy adapted to a pair of Riemannian manifolds. 
Let $(M,g)$ be an $m$ dimensional Riemannian manifold what 
we assume to be oriented and compact for technical reasons. Consider the 
associated volume measure $\mu_g\in\Omega^m(M)$ defined by the aid of the 
Hodge operator associated with the orientation and the metric as 
$\mu_g:=*_g1$. Suppose that it is 
normalized i.e. $\int_M\mu_g=1$. If $\ca_g$ denotes the $\sigma$-algebra of 
$\mu_g$-measurable subsets of $M$ then $g$ improves $M$ to a 
{\it Kolmogorov probability measure space} $(M,\ca_g,p_g)$. It is remarkable 
that these probability spaces in fact do not depend on which 
particular normalized-volume metrics $g$ or $h$ they come from. This is 
because by a theorem of Moser \cite{mos} the only diffeomorphism invariant 
of a smooth positive density over $M$ is its volume. Therefore taking two 
different $(M,\ca_g, \mu_g)$ and $(M,\ca_h,\mu_h)$ there exists an 
orientation-preserving diffeomorphism $f:M\rightarrow M$ such that for every 
$A\in\ca_g$ one finds $f(A)\in\ca_h$ and $\int_A\mu_g=\int_{f(A)}\mu_h$. Thus 
switching to another probability space simply corresponds to identify the 
manifold $M$ not with its particular microstate as in (\ref{szetszedes}) but 
with its different one (\ref{fszetszedes}) still belonging to the same manifold 
macrostate $M$ of the abstract ideal gas $X$. 

Keeping in mind this universality of the Riemannian probablity spaces and 
fixing from now on a particular one $(M,\ca_g,\mu_g)$ we can assign a 
meaning to the emerging probabilities $p_g(A):=\int_{A}\mu_g$ 
for every $A\in\ca_g$ as follows. Motivated in a straightforward way by 
interpreting Theorem \ref{chaitin} above as the inherent indistinguishability 
of the points of the continuum we accept that the apparently simple task of 
identifying or localizing a point $x_0\in M$ within $M$ cannot be carried out. 
The best we can do is to introduce the following    
\vspace{0.1in}

\noindent{\bf Assumption--mathematical form.} 
{\it The number $0\leqq p_g(A)\leqq 1$ is the probability that a 
distinguished point satisfies that $x_0\in A\subseteqq M$.}
\vspace{0.1in}
   
\noindent Accepting this interpretation let 
$\{A_i\}_{i=1,\dots,n}$ be a {\it finite covering of $M$ by $\mu_g$-measurable 
almost-disjoint subsets} i.e. $M=\cup_iA_i$ such that 
$A_i\in\ca_g$ thus $0\leqq p_g(A_i)\leqq 1$ exists for all $i=1,\dots,n$ 
and $p_g(A_i\cap A_j)=0$ for all $i\not=j$. This implies that 
$\sum_ip_g(A_i)=1$. Associated with the metric $g$ and the covering 
$\{A_i\}_{i=1,\dots,n}$ one introduces in a natural way 
the {\it approximate Shannon entropy} of $M$ with respect to the covering:
\begin{equation}
S\big(M,g,\{A_i\}_{i=1,\dots,n}\big):=-\sum_{i=1}^np_g(A_i)\log p_g(A_i)\:\:.
\label{shannon}
\end{equation} 
If the information that the point $x_0\in M$ actually satisfies 
$x_0\in A_i$ is interpreted as saying that ``$M$ is in 
its $i^{\rm th}$ state'' then taking into account the interpretation of 
the probabilities involved we can say that (\ref{shannon}) describes 
the entropy of a ``state'' of $M$ which is the mixture of the ``pure 
states'' $i=1,\dots,n$ with corresponding probabilities $p_g(A_i)$. 
Observe that this formally agrees with the general definition of the 
entropy of a system in statistical physics. It is clear that the 
``knowledge'' about the ``point distribution'' of $M$ is improved if the 
covering is refined. Therefore it is challenging to define the 
entropy of $M$ by taking the limit of (\ref{shannon}) over all coverings, 
if exists. However we cannot expect to come up with any reasonable number in 
this way since taking for example an equipartition i.e. for which 
$p_g(A_i)=\frac{1}{n}$ for all $i=1,\dots,n$ with corresponding entropy 
(\ref{shannon}) then
\[\lim\limits_{n\rightarrow+\infty}
S\big(M,g,n\big)=\lim\limits_{n\rightarrow+\infty}\log n=+\infty\] 
demonstrating that the naive Shannon entropy of the continuum diverges 
at least logarithmically. Nevertheless using the physical language this 
equipartition corresponds to ergodicity of the equilbrium dynamics of the 
continuum in its manifold-macrostate $M$ provided by its orientation-preserving 
diffeomorphisms; hence comparing this formula with the formal entropy 
expression above which also expresses entropy in an equilibrium state under 
ergodic dynamics we find that $\Gamma({\rm Diff}^+(M))\sim n$ as 
$n\rightarrow+\infty$ hence indeed regularization needed. 

However it turns out that this is the only sort of divergence and one can 
renormalize the entropy of a compact Riemannian manifold essentially 
by removing a single logarithmically divergent universal term from 
(\ref{shannon}). To this end we will 
follow \cite[Section 8.3]{cov-tho}. Assume that in $\{A_i\}_{i=1,\dots,n}$ 
every $A_i$ has the form of a closed $m$-ball hence we can 
choose a local coordinate system $(A_i,x^1,\dots, x^m)$ in each of them 
such that there exists a uniform number 
\[\Delta:=\int\limits_{A_i}\dd x^1\dots\dd x^m\] 
for all $i=1,\dots, n$ satisfying $0<\Delta<+\infty$ taking into account the 
orientability of $M$. 
Moreover $p_g(A_i)=\int_{A_i}\mu_g=\int_{A_i}\sqrt{\det g(x^1,\dots,x^m)}
\dd x^1\dots\dd x^m$ therefore introducing the smooth strictly positive 
local function $\rho_i:=\sqrt{\det (g\vert_{A_i})}$ 
by the mean value theorem there exists a point $y_i\in A_i$ such that 
$p_g(A_i)=\rho_i(y_i)\Delta$. Thus (\ref{shannon}) takes the shape  
\[S\big(M,g,\Delta\big)=-\sum\limits_{i=1}^n\rho_i(y_i)\Delta\:
\log\big(\rho_i(y_i)\Delta\big)\]
which is however a highly coordinate-depending expression. To overcome this 
difficulty introduce another Riemannian structure $(M,h)$ having 
normalized volume too i.e. $\int_M\mu_h=1$ with corresponding 
strictly positive local density functions 
$\sigma_j:=\sqrt{\det (h\vert_{A_j})}$ hence  
$p_h(A_j)=\sigma_j(z_j)\Delta$ with some point $z_j\in A_j$. Then the 
Shannon entropy can be expanded like  
\begin{eqnarray}
S\big(M,g,\Delta\big)&=&-\sum\limits_{i=1}^n\rho_i(y_i)\Delta
\:\log\left(\frac{\rho_i(y_i)}{\sigma_i(z_i)}\sigma_i(z_i)\Delta\right)
\nonumber\\
&=&-\sum\limits_{i=1}^n\rho_i(y_i)\Delta\left(
\log\frac{\rho_i(y_i)}{\sigma_i(z_i)}+
\log\sigma_i(z_i)+\log\Delta\right)\nonumber\\
&=&-\sum\limits_{i=1}^n
\log\left(\frac{\rho_i(y_i)}{\sigma_i(z_i)}\right)\rho_i(y_i)\Delta
-\sum\limits_{i=1}^n\log(\sigma_i(z_i))\rho_i(y_i)\Delta -
\log\Delta\sum\limits_{i=1}^n\rho_i(y_i)\Delta\:\:.\nonumber
\end{eqnarray}
Let us say that the countable sequence $\{A_1\}, \{A_1,A_2\},...,
\{A_i\}_{i=1,\dots,n},\dots$ is 
a {\it refinement} (of a finite covering of $M$ by $\mu_g$-measurable
almost-disjoint subsets as above) if for every points $x,y\in M$ with 
$x\not=y$ there exists an $n_{x,y}$ such that for every $n>n_{x,y}$ 
the corresponding $\{A_i\}_{i=1,\dots,n}$ in this sequence contains no single 
element $A_j$ satisfying $x,y\in A_j$. Applying this to our covering with 
uniform balls, refinement implies $\Delta\rightarrow 0$ but not the other way 
round. We make now the following four observations. The first and most 
important is that the ratios of the local functions already extend globally: 
using the globally existing volume-forms $\mu_g,\mu_h\in\Omega^m(M)$ there 
exists a positive smooth function $f:M\rightarrow\R$ satisfying $\mu_g=f\mu_h$ 
and obviously $f\vert_{A_i}=\frac{\rho_i}{\sigma_i}$. We can write this fact as 
$\frac{\rho_i}{\sigma_i}=\frac{\dd\mu_g}{\dd\mu_h}\big\vert_{A_i}$ in terms 
of the globally well-defined Radon--Nikodym derivative of the involved 
measures. The second observation is that 
$\sum_i\rho_i(y_i)\Delta=\int_M\mu_g=1$. 
These two observations make sure that the first term on the right 
hand side converges to a coordinate-free i.e. globaly well-defined integral 
$-\int_M\log\Big(\frac{\dd\mu_g}{\dd\mu_h}\Big)\mu_g$ during a refinement. 
Thirdly, the second term on the right hand side is a coordinate-dependent 
hence not well-defined number $I(M,g,h,\Delta)$ nevertheless satisfying 
$\vert I(M,g,h,\Delta)\vert\leqq c(M,h)\int_M\mu_g=c(M,h)$ hence remains 
bounded (possibly vanishes) during a refinement. Finally the third term is 
equal to $\log\Delta$ representing the already recognized logarithmic 
divergence in the Shannon entropy. 

Putting all of these findings together we arrive at the 
following result which can be understood as the appropriate 
renormalization of the Shannon entropy of a compact Riemannian space; that 
is upon removing two ill-defined terms from it we come up with a 
well-defined i.e. diffeomorphism-invariant quantity:

\begin{theorem}{\rm (cf. \cite[Theorem 8.3.1]{cov-tho})} Let $(M,g)$ be an 
$m$ dimensional compact oriented Riemannian manifold having unit volume 
and $\{A_i\}_{i=1,\dots,n}$ as above which is uniform in the sense that 
there exists a local coordinate system 
$(A_i,x^1,\dots,x^m)$ such that $\Delta=\int_{A_i}\dd x^1\dots\dd x^m$ is a 
positive number independent of $i=1,\dots,n$. Let $(M,h)$ be another 
Riemannian structure having unit volume too. 

Then the approximate Shannon entropy $S(M,g,\Delta)$ under the refinement of 
the corresponding covering behaves like 
\[\lim\limits_{\Delta\rightarrow 0}
\big(S(M,g,\Delta)+I(M,g,h,\Delta)+\log\Delta\big)=
-\int\limits_M\log\left(\frac{\dd\mu_g}{\dd\mu_h}\right)\mu_g\]
which means that the sum of three expressions including the 
approximate Shannon antropy and which are ill-defined separately in 
different ways, already converge under refinements to a well-defined expression.

The quantity $-\int_M\log\Big(\frac{\dd\mu_g}{\dd\mu_h}\Big)\mu_g$ is called 
the {\rm Kullback--Leibler relative entropy} of $(M,g)$ with respect to 
$(M,h)$ regarded as an ambient fixed Riemannian structure.
\hspace{2.6in}$\square$
\label{kl}
\end{theorem} 

\begin{remark}\rm The Kullback--Leibler relative entropy is 
invariant under orientation-preserving diffeomorphisms of $M$ however is 
{\it not} symmetric under $g\leftrightarrow h$; in particular 
it follows from the Jensen inequality that it is always non-negative and is 
equal to zero if and only if $\mu_g$-almost everywhere $\mu_g=\mu_h$ holds 
\cite[Theorem 8.6.1]{cov-tho}. This is the case for instance if $(M,g)$ 
and $(M,h)$ are isometric. Moreover recalling again Moser's 
theorem \cite{mos} without loss of generality we can assume that for instance 
$\mu_h$ is equal to a once and for all fixed density $\mu_0\in\Omega^m(M)$ 
having unit volume. Therefore the Kullback--Leibler relative entropy 
does not really depend on $(M,h)$ hence it can be understood as a quantity 
which measures the ``knowledge'' on various changing geometries like $(M,g)$ 
from the ``viewpoint'' of a once and for all fixed but otherwise arbitrary 
geometry like $(M,h)$.
\end{remark}

\noindent The Kullback--Leibler relative entropy admits a hypersurface 
formulation too:

\begin{theorem} {\rm (An analogue of Hawking's area theorem 
\cite{haw1})} Let $M$ be an $m>1$ dimensional compact oriented manifold 
with non-empty connected boundary $\partial M$ and let $(M,g_i)$ with $i=0,1$ 
be two smooth Riemannian structures on it having (non-normalized) 
volume-forms $\mu_i\in\Omega^m(M)$ and corresponding 
volumes $0<V_i<+\infty$ respectively.

Then, upon modifying the metric $g_0$ with an inessential homothety if 
necessary, an equality  
\[-\frac{1}{V_1}\int\limits_M\log\left(\frac{\dd\mu_1}{\dd\mu_0}\right)\mu_1+
\log\frac{V_1}{V_0}={\rm Area}_1(\partial M)-{\rm Area}_0(\partial M)\]
holds where ${\rm Area}_i(\partial M)=\int_{\partial M}\sigma_i$ is the area of 
the boundary with respect to its induced orientation and the 
surface-form $\sigma_i$ provided by the volume-form $\mu_i$. 

Therefore taking into account the non-negativity of the left hand side as well, 
an inequality 
\[{\rm Area}_1(\partial M)\geqq{\rm Area}_0(\partial M)\]
exists. Referring to our interpretation of the Kullback--Leibler relative 
entropy above, the surface area of the boundary with respect to the ``unknown'' 
geometry $(M,g_1)$ is not smaller than the surface area of the boundary 
with respect to the ``known'' (fixed but modified with a homothety if 
necessary) geometry $(M,g_0)$.
\label{hawking}
\end{theorem}

\begin{remark}\rm 1. Before embarking upon the proof let us recall that a 
{\it homothety} is the scaling of a Riemannian metric with a 
positive constant i.e. $g_0\mapsto c^2g_0$ with an arbitrary $0\not=c\in\R$. 
Such a constant scaling is generally considered (by both mathematicians 
and physicists) as irrelevant. At this level 
of generality an application of a homothety on $(M,g_0)$ might be necessary 
in order our statements to be valid, see the proof below. However it is 
possible that in more restricted situations (like being $(M,g_1)$ a 
spatial section ``preceded by'' $(M,g_0)$ in a common ambient space-time 
satisfying the Einstein equation, etc.) performing homotheties turns out to 
be unnecessary. 

2. Moreover with some technical effort the theorem in an appropriate form 
could be stated over non-compact manifolds as well however we skip that 
formulation here. 
\end{remark}

\begin{proof} Expand the relative entropy expression like 
\begin{eqnarray}
-\int\limits_M\log\left(\frac{\dd\mu_1}{\dd\mu_0}\frac{V_0}{V_1}
\right)\frac{\mu_1}{V_1}&=&{\rm Area}_1(\partial M)
-\int\limits_M\log\left(\frac{\dd\mu_1}{\dd\mu_0}
\frac{V_0}{V_1}\right)\frac{\mu_1}{V_1}-{\rm Area}_1(\partial M)\nonumber\\
&=&{\rm Area}_1(\partial M)-
\left(\:\:\int\limits_M\log\left(\frac{\dd\mu_1}{\dd\mu_0}
\frac{V_0}{V_1}\right)\frac{\mu_1}{V_1}+
\int\limits_M{\rm Area}_1(\partial M)\frac{\mu_1}{V_1}\right)\nonumber\\
&=&{\rm Area}_1(\partial M)-
\int\limits_M\log\left(\frac{\dd\mu_1}{\dd\mu_0}\frac{V_0}{V_1}
{\rm e}^{{\rm Area}_1(\partial M)}
\right)\frac{\dd\mu_1}{\dd\mu_0}\frac{V_0}{V_1}\frac{\mu_0}{V_0}\nonumber
\end{eqnarray}
and consider the following Dirichlet problem:
\[\left\{\begin{array}{lll}
\Delta_0u&=&\frac{\dd\mu_1}{\dd\mu_0}\frac{V_0}{V_1}
\log\left(\frac{\dd\mu_1}{\dd\mu_0}\frac{V_0}{V_1}\:
{\rm e}^{{\rm Area}_1(\partial M)}\right)\\
u\vert_{\partial M}&=&0\end{array}\right.\]
where $\Delta_0$ is the scalar Laplacian on 
$(M,g_0)$. It is known (cf. e.g. \cite[Section 5.1]{tay}) that 
this problem has a unique solution $u\in C^\infty(M;\R)$ whose smoothness 
follows from that of $g_0$ and the inhomogeneous term on the right hand side. 
We proceed further by applying the divergence expression ${\rm div}X=L_X\mu_0$ 
where $L_X$ is the Lie derivative along a vector field $X\in C^\infty(M;TM)$ 
and then Cartan's magic formula $L_X\omega=\dd\omega(X\:,\:\cdot\:)+
\dd(\omega(X,\:\cdot\:))$ and finally Stokes' theorem to get  
\begin{eqnarray}
\int\limits_M\frac{\dd\mu_1}{\dd\mu_0}\frac{V_0}{V_1}
\log\left(\frac{\dd\mu_1}{\dd\mu_0}\frac{V_0}{V_1}
{\rm e}^{{\rm Area}_1(\partial M)}\right)
\frac{\mu_0}{V_0}&=&\frac{1}{V_0}\int\limits_M(\Delta_0u)\mu_0=
\frac{1}{V_0}\int\limits_M{\rm div}({\rm grad}\:u)\mu_0=
\frac{1}{V_0}\int\limits_ML_{{\rm grad}\:u}\:\mu_0\nonumber\\
&=&\frac{1}{V_0}\int\limits_M\!\!\!\big(\dd\mu_0({\rm grad}\:u,\cdot)+
\dd (\mu_0({\rm grad}\:u,\cdot))\big)=
\frac{1}{V_0}\int\limits_{\partial M}\!\!\!\mu_0({\rm grad}\:u,\cdot)\nonumber\\
&=&\frac{1}{V_0}\int\limits_{\partial M}g_0(N_0\:,\:{\rm grad}\:u)\sigma_0
\nonumber
\end{eqnarray}
where $N_0$ denotes the unit normal to $\partial M$ with respect to the 
orientation of $M$ and the metric $g_0$. 

Consider the function 
$g_0(N_0\:,\:{\rm grad}\:u):\partial M\rightarrow\R$. 
Because $u$ is surely not constant over $M$ but is surely 
constant along $\partial M$ we know that ${\rm grad}\:u\not=0$ and 
is parallel with $N_0$ hence $g_0(N_0\:,\:{\rm grad}\:u)$ is a 
not-identically-zero function. Let $Y\in C^\infty (\partial M;T(\partial M))$ 
be a tangent field. If $\nabla$ denotes the Levi--Civita 
connection of $g_0$ over $M$ then 
$Yg_0(N_0\:,\:{\rm grad}\:u)=g_0(\nabla_Y N_0\:,\:{\rm grad}\:u)+g_0(N_0\:,
\:\nabla_Y{\rm grad}\:u)$. 
However by the definitions of $Y$ and $N_0$ they are not only orthogonal 
but even $\nabla_YN_0=0$ holds; moreover we also find that 
$\nabla_Y{\rm grad}\:u={\rm grad}(Yu)=0$ because $u=0$ along the boundary. 
Therefore we conclude that $Yg_0(N_0\:,\:{\rm grad}\:u)=0$ for every tangent 
field hence by the connectivity of $\partial M$ in fact 
$g_0(N_0\:,\:{\rm grad}\:u)=a$ is a 
non-zero constant (depending on $g_1$ through $u$ as well). 

We want to eliminate this constant upon applying a homothety 
with $0\not=c\in\R$ on the metric $g_0$. Beyond 
the scaling $g_0\mapsto c^2g_0$ of the metric itself 
let us collect the induced scalings of the other things involved in the last 
integral too. These are $N_0\mapsto c^{-1}N_0$ and $\mu_0\mapsto 
c^m\mu_0$ hence $V_0\mapsto c^mV_0$ however $\sigma_0\mapsto 
c^{\frac{m-1}{m}}\sigma_0$. Next, the function comprising the 
inhomogeneous term in the Dirichlet problem above is not sensitive for the 
homothety on $g_0$ meanwhile $\Delta_0\mapsto c^{-2}\Delta_0$; therefore 
$u\mapsto c^2u$. In addition to this by definition 
${\rm grad}\:f=g_0(\dd f,\cdot)$ with $g_0$ here being the {\it inverse} metric 
hence scaling as $g_0\mapsto c^{-2}g_0$; thus 
${\rm grad}\mapsto c^{-2}{\rm grad}$ yielding eventually 
${\rm grad}\:u\mapsto{\rm grad}\:u$. Putting all of these together 
the last integral on the right hand side above scales as 
\[\frac{1}{V_0}\int\limits_{\partial M}g_0(N_0\:,\:{\rm grad}\:u)\sigma_0
\longmapsto\frac{1}{c^mV_0}
\int\limits_{\partial M}c^2g_0\big(c^{-1}N_0\:,\:{\rm grad}\:u
\big)c^{\frac{m-1}{m}}\sigma_0=
\frac{c^{1-m}a}{V_0}{\rm Area}_{c^2g_0}(\partial M)\]
demonstrating that if $\dim_\R M=m>1$ then 
we can adjust the homothety so that $\frac{c^{1-m}a}{V_0}=1$ 
rendering the integral in question equal 
to ${\rm Area}_{c^2g_0}(\partial M)$. However the original integral on 
the left hand side is invariant under homotheties. Therefore, upon modifying 
the metric $g_0$ with a homothety $g_0\mapsto c^2g_0$ if necessary, we come 
up with the equality of the theorem. 

The inequality then also follows taking into account the non-negativity 
of the left hand side provided by the aforementioned non-negativity of 
the Kullback--Leibler relative entropy (observe again the asymmetry of 
the relative entropy formula in its metric content!). 
\end{proof} 

\noindent After these preparations we are in a position to offer 
a mathematically meaningful definition of the entropy of the continuum at 
least in a relative way by comparing its two particular microstates within a 
common compact-orientable-manifold-macrostate. Let $M$ be a compact 
orientable $m$-manifold carrying two strictly positive measures 
$\mu_0,\mu_1\in\Omega^m(M)$ satisfying $\int_M\mu_i=V_i$ for $i=0,1$. Using 
$\frac{\mu_0}{V_0}$ as a fixed reference 
measure as so far we can make $\Omega^0(M)$, the space of smooth functions 
over the compact $M$, a Banach space $L^\infty(M,\frac{\mu_0}{V_0})$ by 
completing it with respect to the norm $\Vert f\Vert_{L^\infty}:=
\mbox{$\frac{\mu_0}{V_0}$-${\rm ess\:sup}_{x\in M}\vert f(x)\vert$}$ for 
every $f\in\Omega^0(M)$. The dual space of $L^\infty(M,\frac{\mu_0}{V_0})$ is 
$L^1(M,\frac{\mu_0}{V_0})$ and contains $\Omega^m(M)$ because the formula 
$F_\omega (f):=\int_Mf\omega$ by extension gives rise to a continuous linear 
functional on $L^\infty(M,\frac{\mu_0}{V_0})$ for every 
$\omega\in\Omega^m(M)$. The norm on 
$\Omega^m(M)\subset L^1(M,\frac{\mu_0}{V_0})$ is $\Vert\omega\Vert_{L^1}=
\int_M\vert\omega\vert=\int_M\vert\frac{\dd\omega}
{\dd\mu_0}\vert\mu_0$ and it follows that $\frac{\mu_1}{V_1}$ belongs to 
the unit ball in the dual space $L^1(M,\frac{\mu_0}{V_0})$. Introducing the 
${\rm weak}^*$-topology on $L^1(M,\frac{\mu_0}{V_0})$ generated by the 
seminorms $\Phi_f$ of the form 
$\Phi_f(\omega):=\vert\int_Mf\omega\vert$ it is easy to check that 
$\frac{\mu_1}{V_1}\mapsto-\int_M\log\Big(\frac{\dd\mu_1}{\dd\mu_0}
\frac{V_0}{V_1}\Big)\frac{\mu_1}{V_1}$
is continuous. But by Alaoglu's theorem (e.g. \cite[p. 484]{tay}) the unit 
ball in $L^1(M,\frac{\mu_0}{V_0})$ is compact in 
the ${\rm weak}^*$-topology consequently this map attains its maximum 
somewhere hence 
\[S(M):=\sup\limits_{\mu_1\in\Omega^m(M)}\:\:-\int\limits_M
\log\left(\frac{\dd\mu_1}{\dd\mu_0}\frac{V_0}{V_1}\right)\frac{\mu_1}{V_1}\]
is a finite number. It is independent of the choice for
the reference measure $\frac{\mu_0}{V_0}$ too by the aid of its diffeomorphism 
invariance and Moser's theorem \cite{mos}. Moreover it satisfies the 
(sub)additivity $S(M\#N)\leqq S(M)+S(N)$ and $S(M\times N)=S(M)+S(N)$ 
under taking connected sum or Descartes product, respectively. Thus 
collecting all of our (including interpretational, with special attention to 
{\bf Assumption--mathematical form}) efforts so far we put 
\begin{definition}
Let $X$ be an abstract set whose cardinality coincides with 
that of the continuum in ZFC set theory. Then 
\begin{equation}
\mbox{\rm Entropy of the set $X$ in its compact-orientable-manifold-macrostate 
$M:=S(M)$}
\label{entropia1}
\end{equation}
where $0<S(M)<+\infty$ is the quantity introduced above.
\end{definition}

\noindent In the case of a manifold-with-boundary we can compare 
(\ref{entropia1}) with Theorem \ref{hawking} to obtain a two-sided inequality
\begin{equation}
0\leqq{\rm Area}_1(\partial M)-{\rm Area}_0(\partial M)\leqq S(M)
\label{egyenlotlensegek}
\end{equation}
for an arbitrary pair $(M,g_0)$ and $(M,g_1)$ (upon applying a homothery on the 
former member).

Before closing this section let us also return to the problem of the volume of 
the diffeomorphism group here for a moment; comparing our entropy 
definitions so far we come up with 
\[\Gamma ({\rm Diff}^+(M)):={\rm e}^{S(M)}<+\infty\]
as a reasonable choice at least in the compact orientable setting. 

To summarize, we have sketched a framework in which the {\it inherent 
uncertainty} or {\it fuzziness} of the arithmetical continuum i.e. the 
set $\R$ of real numbers or more 
generally any differentiable manifold, can be interpreted as a {\it 
non-zero entropy} of the arithmetical continuum (cf. \cite{sch1,sch2}), 
quantitatively captured by (\ref{entropia1}) at least in the compact case.


\section{Secondary Noether theory and the entropy of black holes}
\label{three}


Keeping in mind the results of Section \ref{two} and recalling now 
\cite[Section 3]{ete2} we would like to approach the formula 
(\ref{entropia1}) again but by passing from mathematics to physics. Namely, 
we shall consider classical physical theories 
over physical space-time such that in the mathematical description of these 
theories the physical space-time is modeled on a differentiable manifold 
possessing the property (\ref{szetszedes}) or more generally 
(\ref{fszetszedes}). Then, we shall ask ourselves: does the inherent 
uncertainty or fuzziness 
of the arithmetical continuum, just recognized in the mathematical model of the 
physical theory, ``pop up'' somehow among the physical propositions of the 
physical theory? Putting differently: does this fuzziness somehow ``lift'' 
from the mathematical level to the physical level of the physical theory? 
Since we have found some similarities between this purely mathematical 
uncertainty or fuzziness of the arithmetical continuum and the physical 
concept of {\it entropy}, we are going to seek entropylike phenomena in 
those physical theories which are particularly sensitive for the physical 
structure of space-time. If these sought entropylike phenomena happen to 
have a pure set-theoretical origin introduced by the mathematical 
description of the physical theory, then we expect them to have something 
to do with {\it diffeomorphisms} of the underlying differentiable manifold 
modeling physical space-time; for the expression (\ref{entropia1}) 
is diffeomorphism-invariant hence the entropy it describes is invariant 
under diffeomorphisms. Apart from this, if diffeomorphisms are in addition 
{\it symmetries} of the physical theory we are dealing with then we may as 
well try to identify these entropylike things with Noether charges 
associated with diffeomorphism symmetry.

By {\it Noether's theorem} in a broad physical sense one means 
that ``to every continuous symmetry of a physical theory a 
quantity can be assigned which is conserved''. It may happen however 
that this conserved quantity, the {\it Noether charge}, vanishes. Our goal 
is to demonstrate that even in this trivial case certain non-trivial de Rham 
cohomology classes can still be interpreted as 
``secondary Noether charges'' associated with this symmetry of the theory. 
There is an analogouos situation in algebraic topology. Consider a complex 
vector bundle $E$ over a topological space $X$. Recall that for all 
$i=0,\dots,{\rm rk}\:E$ the $i^{\rm th}$ Chern class of $E$ takes value in 
$H^{2i}(X;\Z)$. 
Therefore, if it happens that $X$ has vanishing {\it even} dimensional 
singular cohomology then {\it characteristic classes} cannot be used to 
distinguish complex vector bundles over it.\footnote{A simple example for 
this failure is provided by complex rank two vector bundles with structure 
group ${\rm SU}(2)$ over the $5$-sphere $S^5$. Then, on the one hand, 
isomorphism classes of these type of vector bundles are classified by the 
group $\pi_4({\rm SU}(2))\cong\pi_4(S^3)\cong\Z_2$ hence there are 
precisely two different such bundles up to isomorphism over $S^5$; 
meanwhile, on the other hand, $H^k(S^5;\Z)\cong\{0\}$ if $k\not=0,5$ hence 
all Chern classes of these bundles are trivial.} However if 
$X$ is a manifold $M$ then one can still introduce the so-called {\it 
secondary} or {\it Chern--Simons characteristic classes} taking values, as 
a cohomological shift, in {\it odd} dimensional cohomology \cite{che-sim}. 
Motivated by this consideration as well as those in \cite{dol,wal3} we 
proceed as follows.
 
For completeness and convenience let us recall how standard Noether 
theory works in case of a classical 
relativistic field theory. We are going to skip all the technical details 
here but emphasize that in case of a closed, orientable Riemannian 
$4$-manifold all of our considerations below are rigorous mathematical 
statements; therefore we have a reason to expect that with appropriate 
technical modifications all the stuff remains valid in physically more 
realistic situations. 

So let $(N,h)$ be a four dimensional (non-)closed oriented 
(pseudo-)Riemannian manifold representing space-time and let $\Phi$ denote 
the full field content of a classical field theory over $(N,h)$ defined by a 
Lagrangian density $L(\Phi,h)\in\Omega^4(N)$. Note that by definition the 
Lagrangian is not a function but a $4$-form over $N$ allowing one to talk about 
the corresponding action $S(\Phi ,h)=\int_NL(\Phi,h)$ defined by 
integration over $N$. Let $\cm$ be the configuration space 
of {\it all} (but belonging to a nice function class) $(\Phi,h)$-field 
configurations over $N$ i.e. its elements are {\it not} identified by 
diffeomorphisms, gauge, etc. transformations. Consider a 
differentiable curve $C:\R\rightarrow \cm$. We say that $C$ is a {\it symmetry} 
of the theory $L(\Phi,h)$ if its action $S(\Phi,h)$ is constant along $C$ 
that is, $S(C(t))=S(\Phi (t),h(t))=$const. for all $t\in\R$. 
Writing $(\Phi,h):=(\Phi (0),h(0))$ and using physicists' usual 
notation define ``the infinitesimal variation of the action at 
$(\Phi,h)$ along $C$'' by   
\[\delta_CS(\Phi,h):=\lim\limits_{t\rightarrow 0}
\frac{1}{t}(S(C(t))-S(C(0)))=\int\limits_N\lim\limits_{t\rightarrow 0}
\frac{1}{t}(L(C(t))-L(C(0)))=:\int\limits_N\delta_CL(\Phi,h)\] 
where $\delta_CL(\Phi,h)\in\Omega^4(N)$ is the ``infinitesimal variation 
of the Lagrangian at $(\Phi,h)$ along $C$''.\footnote{Using standard notations 
of differential geometry if $\dot{C}$ is the derivative of $C$ at $t=0$ 
then $\delta_CL=\dot{C}(L):\cm\rightarrow\Omega^4(N)$. By specializing the 
variation further we can demand $\delta^2=0$ hence we can formally treat 
$\delta$ as an exterior derivative on the infinite dimensional manifold 
$\cm$ \cite[and references therein]{dol} and can introduce the 
$\Omega^4(N)$-valued $1$-form $\delta L$ on $\cm$. Then $\delta_CL=\delta L
(\dot{C}):\cm\rightarrow\Omega^4(N)$ as well.} 
Assume that $(N,h)$ is smooth and let $\Delta_h:=\dd\dd^*+\dd^*\dd$ denote 
its Hodge Laplacian; recall \cite[Chapter 5]{tay} that if 
$\omega\in\Omega^4(N)$ the partial 
differential equation $\Delta_h\varphi =\omega$ has a smooth solution 
$\varphi$ if and only if $\int_N\omega=0$. By definition of a symmetry 
$\int_N\delta_CL(\Phi,h)=0$ hence there exists an element 
$\varphi_C\in\Omega^4(N)$ satisfying $\Delta_h\varphi_C =\delta_CL(\Phi,h)$. 
However $\Delta_h\varphi_C=\dd\dd^*\varphi_C +\dd^*\dd\varphi_C =
\dd\dd^*\varphi_C$ consequently picking any $\eta_C\in\Omega^2(N)$ and 
putting 
\begin{equation}
\theta_C:=\dd^*\varphi_C+\dd\eta_C
\label{teta}
\end{equation}
we succeeded to find an element $\theta_C\in\Omega^3(N)$ such that 
$\dd\theta_C=\delta_CL(\Phi,h)$. Note first that, although 
$\varphi_C$ is well-defined only up to a harmonic $4$-form i.e. an element 
$\varphi\in{\rm ker}\:\Delta_h$, the $3$-form $\theta_C$ is not sensitive 
for this ambiguity because taking into account its harmonicity, 
the general solution $\varphi$ above is both closed ($\dd\varphi=0$) and 
co-closed ($\dd^*\varphi= 0$) hence $\theta_C=\dd^*\varphi_C+\dd\eta_C=
\dd^*(\varphi_C +\varphi)+\dd\eta_C$. Secondly, the ``gauge freedom'' i.e. the 
$\eta_C$-ambiguity can be fixed as well by imposing the ``Coulomb gauge 
condition'' $\dd^*\theta_C=0$. Indeed, 
$\dd^*\theta_C={\dd^*}^2\varphi_C+\dd^*\dd\eta_C=\dd^*\dd\eta_C=0$ 
(together with the Hodge decomposition theorem) implies $\dd\eta_C=0$. 
Therefore, given a symmetry $C$ of the theory we come up with a 
$\theta_C\in\Omega^3(N)$ which satisfies
\[\left\{\begin{array}{lll}
\dd\theta_C&=&\delta_CL(\Phi,h)\\
\dd^*\theta_C&=&0\\
\int_N\dd\theta_C&=&0
\end{array}\right.\]
along $N$ and this $3$-form is well-defined in the sense that 
it is unique and depends only on the symmetry represented by the curve $C$ as 
expected.\footnote{If $(\Phi_0,h_0)\in\cm$ is a {\it critical point} of the 
action then $\delta_CS(\Phi_0,h_0)=\int_N\delta_CL(\Phi_0,h_0)=
\int_N\delta L(\Phi_0,h_0)(\dot{C})=0$ for {\it all} curves 
passing through the critical point. Hence 
$\int_N\delta L(\Phi_0,h_0)(\dot{C})=0$ for all $\dot{C}\not=0$ thus 
in fact $\delta L(\Phi_0,h_0)=0$ which is the resulting {\it field 
equation} of the theory.}

Proceeding further, we call the Hodge dual $1$-form  
$j_C:=*_h\theta_C\in \Omega^1(N)$ the {\it Noether current} associated with the 
symmery $C$ moreover for a (spacelike) 
hypersurface-without-boundary $S\subset N$ put $q_{C,S}:=\int_S*_hj_C$ and 
call it the {\it Noether charge} associated with the symmetry $C$. The 
Noether charge satisfies 
\[q_{C,S_1}-q_{C,S_2}=\pm\!\!\!\!\!\int\limits_{W(S_1,S_2)}\dd\theta_C =0\]
by applying Stokes' theorem on a domain $W(S_1,S_2)\subseteqq N$ with 
induced orientation and oriented boundary $\partial W(S_1,S_2)=
S_1\sqcup (-S_2)$. (Here we strictly speaking assume that the 
variation vanishes on the complementum $N\setminus W(S_1,S_2)$ hence 
$\int_{W(S_1,S_2)}\dd\theta_C=\int_N\dd\theta_C=0$ indeed.) Consequently the 
real number $q_C:=q_{C,S}$ is a well-defined 
conserved (i.e. independent of the spacelike surface $S$) 
quantity associated with the symmetry of the theory in this sense. 

Thanks to the gauge fixing condition $\dd^*\theta_C=0$ the Noether current 
looks like $j_C=*_h\theta_C=\pm\dd*_h\varphi_C$ consequently 
$j_C\in [0]\in H^1(N)$ i.e. the current represents the trivial cohomology 
class in the first de Rham cohomology. We may then ask ourselves what 
about $\theta_C$ from the de Rham theoretic viewpoint? 
Does it represent a cohomology class in $H^3(N)$? Still working in 
the gauge $\dd^*\theta_C=0$, assume 
$\dd\theta_C=0$ holds; then via (\ref{teta}) we get $\Delta_h\varphi_C=0$ 
implying $\varphi_C$ is harmonic hence $\theta_C=\dd^*\varphi_C =0$. 
Therefore we find that $\dd\theta_C=0$ if and only if $\theta_C=0$. 
Consequently $\theta_C\in\Omega^3(N)$ represents a 
cohomology class $[\theta_C]\in H^3(N)$ if and only if $\theta_C=0$ and the 
associated Noether charge $q_C=\int_S*_hj_C=\pm\int_S\theta_C=0$ is 
trivial rendering the classical Noether theory useless in this situation. 

Let us focus attention to this trivial case i.e. when for a symmetry 
$C$ of the theory $L(\Phi,h)$ the associated total derivative satisfies 
$0=[\theta_C]\in H^3(N)$ (by exploiting the gauge fixing condition 
$\dd^*\theta_C=0$ too). The gauge fixing also implies $\dd^*\varphi_C=0$ 
as we have seen hence the general expression (\ref{teta}) reduces to 
\[0=\dd\eta_C\] 
saying that $\eta_C$ itself represents a cohomology class in $H^2(N)$. 
Consequently in this situation---which is trivial from the variational 
viewpoint in the sense that it yields vanishing primary 
Noether theory, but not trivial from the topological viewpoint in the sense 
that $H^2(N)\not\cong\{0\}$ may hold---we can still introduce a 
{\it secondary} or {\it topological Noether current} $J_C
\in\Omega^2(N)$ by putting $J_C:=*_h\eta_C$. Then taking any two dimensional 
submanifold-without-boundary 
$\Sigma\subset N$ the corresponding {\it secondary} or {\it topological 
Noether charge} $Q_{C,\Sigma}:=\int_\Sigma *_hJ_C$ is well-defined in 
the sense that it depends only on the chosen de Rham cohomology class 
$[*_hJ_C]\in H^2(N)$. Moreover 
\[Q_{C,\Sigma_1}-Q_{C,\Sigma_2}=\pm\!\!\!\!\!
\int\limits_{W(\Sigma_1,\Sigma_2)}\dd\eta_C=0\]
by Stokes' theorem as before. (This time 
$W(\Sigma_1,\Sigma_2)\subset N$ is a sub-$3$-manifold with induced orientation 
and oriented boundary $\partial W(\Sigma_1,\Sigma_2)=
\Sigma_1\sqcup (-\Sigma_2)$.) Consequently we have a conserved quantity 
in the sense that the number $Q_{C,\Sigma}=:Q_{C,[\Sigma]}$ depends only on 
$[*_hJ_C]\in H^2(N)$ and the singular homology class $[\Sigma ]\in H_2(N;\Z )$. 
Although it is not necessary, just for aesthetical reasons we can suppose 
without loss of generality that $\eta_C$ is the unique harmonic represnentative 
of $[\eta_C]$ hence both $\eta_C$ and $J_C=*_h\eta_C$ are closed that is, 
represent cohomology classes within $H^2(N)$.  

Note that, regardless what the symmetry $C$ actually is, in order 
$Q_{C,[\Sigma]}$ not to be trivial i.e., 
$Q_{C,[\Sigma]}\not=0$, we need $[0]\not=[\Sigma ]\in H_2(N;\Z )$ 
as well as $[0]\not=[*_hJ_C]\in H^2(N)$. Both conditions are met if 
we demand $N$ to satisfy the topological condition that the free 
part of its second singular homology group $H_2(N;\Z)_{\rm free}\cong
\Z^{{\rm rk}\:H_2(N;\Z)}$ be non-zero (i.e., the rank of 
$H_2(N;\Z)$ be non-zero). Moreover, at this level of generality all 
cohomology classes in $H^2(N)$ are permitted to play the role 
of $[*_hJ_C]$ consequently the number of linearly independent secondary Noether 
currents is equal to $b^2(N)$. Therefore if ${\rm rk}\:H_2(N;\Z)>1$ that is, 
$b^2(N)>1$ then we have ``too many'' options to introduce non-trivial 
secondary Noether charges for a given symmetry. 
Consequently, the optimal situation for this secondary theory is 
when ${\rm rk}\:H_2(N;\Z)=1$ (at this level of generality). 
  
To summarize, we have proved:

\begin{lemma} {\rm (cf. \cite[Lemma 3.1]{ete2})} Let $(N,h)$ be a (non-)closed 
oriented (pseudo-)Riemannian $4$-manifold satisfying the topological condition 
$H_2(N;\Z)_{\rm free}\not\cong\{0\}$. Let moreover a 
classical relativistic field theory be given over $(N,h)$ defined 
by its Lagrangian density $L(\Phi,h)\in\Omega^4(N)$. Assume that 
$C:\R\rightarrow\cm$ is a symmetry of the theory such that the corresponding 
total derivative $\theta_C\in\Omega^3(N)$ satisfying the gauge fixing 
condition $\dd^*\theta_C=0$ is closed i.e. $\dd\theta_C=0$ 
(hence $\theta_C=0$ therefore in fact $[\theta_C]=0\in H^3(N)$). 

Then there exist a $2$-form $0\not=J_C\in \Omega^2(N)$ 
representing a non-trivial de Rham 
cohomology class $0\not=[*_hJ_C]\in H^2(N)$ as well as a closed 
oriented surface $\Sigma\subset N$ representing a non-trivial 
singular homology class $[0]\not=[\Sigma ]\in H_2(N;\Z)_{\rm free}$ such 
that the associated quantity $Q_{C,\Sigma}:=\int_\Sigma *_hJ_C\in\R$ is not 
zero and depends only on $[*_hJ_C]\in H^2(N)$ and $[\Sigma]\in 
H_2(N;\Z)_{\rm free}$\:. We denote this quantity by $Q_{C,[\Sigma]}$ and call 
the {\rm secondary} or {\rm topological Noether charge} associated with the 
symmetry $C$ and  the classes $[*_hJ_C]$ and $[\Sigma]$. 
For a given symmetry $C$ the number of linearly independent cohomology 
classes $[*_hJ_C]$ is equal to ${\rm rk}\:H_2(N;\Z)$.\hspace{5.1in}$\square$
\label{noether}
\end{lemma} 

\noindent Note that what we have done is in fact simple: We interpret the 
{\it a priori} existing cohomology classes of $N$ as certain physical 
quantities whenever a theory $L(\Phi,h)$, possessing certain type of 
symmetries, has been formulated over $N$.

Let us apply this theory for diffeomorphisms in 
pure gravity in four dimensions. Let $\nu_h\in\Omega^4(N)$ defined by 
$\nu_h:=*_h1$ be the volume-form of $(N,h)$ and take the usual 
Einstein--Hilbert Lagrangian 
$L_{EH}(h):=({\rm Scal}_h-2\Lambda )\nu_h$ with cosmological constant 
$\Lambda\in\R$ and consider its variation with respect to a $1$-parameter 
subgroup $\{f_t\}_{t\in\R}$ of the orientation-preserving diffeomorphism 
group ${\rm Diff}^+(N)$ of the underlying space-time manifold $N$ 
while the metric $h$ is kept fixed. That is we define our curve by 
$C(t):=f^*_th\in\cm$ for all $t\in\R$. The infinitesimal generator of 
this $1$-parameter subgroup is a compactly supported vector field 
$X\in C^\infty_c(N;TN)$. 
Since a diffeomorphism acts on $k$-forms via pullback and the {\it scalar} 
curvature is invariant under diffeomorphisms, 
$L_{EH}(C(t))=f^*_tL_{EH}(f^*_th)=f^*_tL_{EH}(h)$ hence 
the corresponding infinitesimal variation takes the shape 
\[\delta_CL_{EH}(h)=\lim\limits_{t\rightarrow 0}\frac{1}{t}(1-f^*_t)
L_{EH}(h)=L_X(L_{EH}(h))\]
where $L_X$ denotes the Lie derivative with respect to $X$. 
Substituting the Lagrangian and applying Cartan's formula we thus get 
\begin{eqnarray}
\delta_CL_{EH}(h)&=&\big(\dd({\rm Scal}_h-2\Lambda )\nu_h\big)(X,\:\cdot)+
\dd\big(({\rm Scal}_h-2\Lambda )\nu_h(X,\cdot)\big)\nonumber\\
&=&\dd\big(({\rm Scal}_h-2\Lambda )\nu_h(X,\cdot)\big)\:\:\:.\nonumber
\end{eqnarray}
While $\delta_CL_{EH}(h)\not=0$ in general, nevertheless we find that 
$\delta_CS_{EH}(h)=\int_M\dd\big(({\rm Scal}_h-2\Lambda )\nu_h(X,\cdot)\big)=0$ 
by Stokes' theorem hence the Einstein--Hilbert action itself is 
invariant consequently diffeomorphisms are both off or on shell symmetries 
of general relativity with possibly non-vanishing cosmological 
constant. The associated total derivative up to an exact term 
looks like $\theta_C=({\rm Scal}_h-2\Lambda )\nu_h(X,\cdot)$ 
in some gauge probably {\it not} satisfying the condition $\dd^*\theta_C=0$. 

In order not to get lost in the gauge fixing problem assume instead that 
(i) we are {\it on shell} i.e., Einstein's equation ${\rm Ric}_h=\Lambda h$ 
is valid hence $\theta_C=(4\Lambda-2\Lambda)\nu_h(X,\cdot)=
2\Lambda\nu_h(X,\cdot)$ and (ii) 
the cosmological constant vanishes yielding $\theta_C=0$. Consequently {\it 
the diffeomorphism symmetry in on shell pure gravity with vanishing 
cosmological constant has vanishing associated (primary) Noether charge.} 
However substituting $\theta_C=0$ into (\ref{teta}) (and referring to the Hodge 
decomposition theorem) we get $\dd^*\varphi_C=0$ and $\dd\eta_C=0$ consequently 
in this physically important situation we can interpret the cohomology classes 
$[\eta_C]\in H^2(N)$ as Hodge duals of currents $J_C$ in secondary Noether 
theory. 

As we stressed even in the formulation of Lemma \ref{noether}, interesting 
secondary Noether theory emerges only if the underlying 
manifold is topologically non-trivial in the sense formulated 
there. At this point, by taking e.g. a survey on known solutions 
\cite{ste-kra-mac-hoe-her}, we make an observation which is 
completely independent of our considerations taken so far---hence in our 
opinion is very interesting!---namely: {\it apparently all explicitly 
known $4$ dimensional black hole solutions in vacuum general 
relativity with vanishing cosmological constant satisfy the topological 
condition formulated in Lemma \ref{noether}}. This intuitively means that 
because of some general reason a black hole is even topologically 
recognizable as a two dimensional ``hole'' in space-time. 
In fact with an appropriate restriction this observation can be 
proved \cite{ete1} and can be considered as a global topological 
counterpart of well-known black hole uniqueness theorems \cite{heu}. In 
accordance with this provable version \cite{ete1} we suppose from now on 
that: $(N,h)$ is a $4$ dimensional solution of the Einstein's equation 
${\rm Ric}_h=0$ and describes a single stationary asymptotically flat black 
hole; hence ${\rm rk}\:H_2(N;\Z)=1$ (which apparently corresponds to the case 
that a ``single'' black hole is present). In this case the homology class of 
the ``instantaneous'' event horizon of the black hole as an (immersed) surface 
$i:\Sigma\looparrowright N$ represents a non-zero element 
$[\Sigma]\in H_2(N;\Z)_{\rm free}\cong\Z$. 

Then we proceed as follows: like the original volume-form 
$\nu_h\in\Omega^4(N)$, the induced $2$ dimensional area-form 
$\sigma_h\in\Omega^2(\Sigma )$ of the ``instantaneous'' 
event horizon $\Sigma$ is closed consequently it represents a 
class $[\sigma_h]\in H^2(\Sigma)$ which is 
not zero since the event horizon has finite area. Then exploiting singleness 
and stationarity, the ``instantaneous'' event horizon $\Sigma$ is connected and 
its area ${\rm Area}_h(\Sigma)=\int_\Sigma\sigma_h$ is constant in time 
consequently we can suppose that the area form is proportional to the Hodge 
dual of the secondary Noether current with a time-independent constant. In 
other words with any choice $J_C\in\Omega^2(M)$ for the Hodge dual of the 
secondary Noether current the $\sigma_h$ satisfies that 
${\rm const.}\:[i^*(*_hJ_C)]=[\sigma_h]\in H^2(\Sigma)\cong\R$ therefore  
\begin{eqnarray}
\mbox{Entropy of the black hole in $(N,h)$}&=&{\rm const.}\:{\rm 
Area}_h(\Sigma)={\rm const.}\!\!\!\int\limits_\Sigma\sigma_h=
{\rm const.}\!\!\!\int\limits_\Sigma*_hJ_C\nonumber\\ 
&=&{\rm const.}\:Q_{C,[\Sigma]}\nonumber
\end{eqnarray}
offering a natural way to normalize $Q_{C,[\Sigma]}$ to be {\it equal} to 
the entropy of the black hole i.e.
\begin{equation}
\mbox{Entropy of the black hole in $(N,h)$}=Q_{C,[\Sigma]}={\rm const.}
\:{\rm Area}_h(\Sigma)\:\:.
\label{entropia2}
\end{equation} 
Accepting this choice of normalization therefore a natural physical 
interpretation of this abstract secondary or topological conserved quantity 
also emerges, namely: {\it if a $4$ dimensional space-time $(N,h)$ is a 
solution of the vacuum Einstein's equation with vanishing cosmological 
constant and describes a single stationary asymptotically flat black hole 
then the secondary Noether charge associated with the orientation-preserving 
diffeomorphism invariance of general relativity is not zero and as a secondary 
conserved quantity is equal to the entropy of the black hole} 
(cf. \cite{dol,wal3}). Thus the point is that in this way black hole 
entropy is conceptionally connected with the invariance of the underlying 
differentiable manifold against its own diffeomorphisms. 

We are now in a position to make our crucial observation: both 
the introduced set-theoretic entropy (\ref{entropia1}) previously 
(as the entropy of an abstract set in its manifold-macrostate) 
and the black hole entropy (\ref{entropia2}) here (as the Noether charge 
for the invariance of a stationary black hole space-time under 
diffeomorphisms) give rise to conserved quantities 
assigned to one and the same process namely the permutation of the points of 
differentiable manifolds by diffeomorphisms. This manifests itself in 
their common diffeomorphism invariance. However the two quantities, namely 
(\ref{entropia1}) and (\ref{entropia2}), even more resemble each other if 
the former is computed over a $3$-manifold-with-boundary $M$, cast in a form 
of differences between areas via Theorem \ref{hawking} yielding 
the two-sided inequality (\ref{egyenlotlensegek}); and then this 
manifold-with-boundary $M$ is inserted as a spacelike submanifold $(M,g_1)$ 
into a space-time $4$-manifold $(N,h)$ containing a (not necessarily 
stationary or single) black hole such that $\partial M=\Sigma$ corresponds to 
the  ``instantaneous'' event horizon whose area ${\rm Area}_h(\Sigma)$ 
is therefore equal to ${\rm Area}_1(\partial M)$. Thus 
(\ref{egyenlotlensegek}) written as 
${\rm Area}_0(\partial M)\leqq{\rm Area}_1(\partial M)\leqq
{\rm Area}_0(\partial M)+S(M)$ qualitatively implies 
\begin{eqnarray}
& &\mbox{Entropy of the black hole in $(N,h)$ with ``instantaneous'' 
event horizon $\partial M$}\nonumber\\
& & \approx{\rm const.}\:+\:\mbox{\rm Entropy of $X$ in its 
manifold-macrostate $M\subset N$}
\label{vegeredmeny}
\end{eqnarray}
where ${\rm const.}\approx{\rm Area}_0(\partial M)$ with respect to some 
reference spacelike submanifold $(M,g_0)\subset (N,h)$ as in 
Section \ref{two} throughout. Thus (\ref{vegeredmeny}) can be regarded like 
the decomposition of the physical black hole entropy as mathematically 
appears in general relativity into the sum of an already recognized pure 
geometric term proportional to ${\rm Area}_0(\partial M)$ and an unexpected 
other term proportional to $S(M)$ describing the inherent set-theoretic 
fuzziness of the mathematical model underlying general relativity. Since the 
right hand side of (\ref{vegeredmeny}) contains the positive 
quantity (\ref{entropia1}) it follows that its left hand side 
(\ref{entropia2}) i.e. the entropy of a black hole cannot decrease 
in accord with Hawking's area theorem \cite{haw1}. 

The time has come to complete the circle of our arguments. On the 
mathematical side, the formal concept of the arithmetical continuum (or 
the set $\R$ of real numbers) contains a tacit uncertainty or 
fuzziness in the sense that the effective identification of the 
arithmetical continuum with its individual disjoint constituents, the {\it 
points}, cannot be carried out (our interpretation of Theorem 
\ref{chaitin}). On the physical side, the nowadays accepted 
mathematical formalization of our intuitive concept of the spatial or 
temporal continuum in terms of the arithmetical continuum lifts the purely 
formal---and concerning its Leibnizian monadistic origin, 
metaphysical---entity, the same {\it point} again, to an ontological level. 
We may then ask ourselves whether 
or not this sort of description of space-time in a mathematical model of a 
physical theory introduces a similar uncertainty or fuzziness into the 
physical theory. Let us formulate our question more carefully. In the 
modern understanding by a {\it physical theory} one means a two-level 
description of a certain class of natural phenomena: the theory possesses 
a {\it syntax} provided by its mathematical core structure and a {\it 
semantics} which is the meaning i.e. interpretation of the bare 
mathematical model in terms of physical concepts. Consider a physical 
theory whose semantics contains a description of space and time (like 
general relativity) and its syntax uses the arithmetical continuum to 
mathematically model the thing which corresponds to the space-time 
continuum at the semantical level of the theory (like general relativity). 
Then we may ask whether or not the uncertainty or fuzziness recognized at the 
syntactical level of the physical theory (introduced by the utilization of 
the arithmetical continuum) shows up at the semantical level of the 
physical theory too. To answer this for a given physical theory, one has 
to search among those physical concepts which describe uncertainty, 
fuzziness, or {\it disorder} at the semantical level and check their 
counterparts at the syntactical level. Of course the basic physical 
concept of this kind is the {\it entropy}.

Therefore, in this context, one can be concerned whether or not entropy 
within classical general relativity, appearing in its semantics in the 
form of black hole entropy \cite{bar-car-haw, haw1}, simply comes from its 
syntax i.e. has a pure mathematical origin only (hence probably not 
corresponding to any ``objective'' thing in the world)? This suspicion is 
also supported in some extent by the several controversial (geometrical, 
thermodynamical, quantum and information theoretic, etc.) properties of 
one and the same thing, the black hole entropy. Our analysis of the black hole 
entropy formula in general relativity culminating in its formal factorization 
(\ref{vegeredmeny}) into the sum of a geometric term and a pure 
set-theoretic term points at least {\it in part} towards a set-theoretic 
origin. That is, even if the geometric term in (\ref{vegeredmeny}) indeed 
corresponds to the ``physical part'' of black hole entropy, the next term 
could be a ``pure mathematical'' or more precisely a ``pure set-theoretic'' 
contribution only. This could be an example how certain physical 
statements within the physical theory of general relativity 
get ``contaminated'' by the underlying mathematical model akin to the situation 
in quantum field theory \cite{bae}.

However, beyond the ``balanced'' interpretation above, if one prefers 
one can read (\ref{vegeredmeny}) in two extreme ways as well, going into 
exactly the opposite directions as follows. The first is 
that black hole entropy is of pure mathematical origin without any 
physical content hence e.g. the long-sought as well as quite problematic 
physical degrees of freedom responsible for black hole entropy (far from being 
complete cf. e.g. \cite{str-vaf}) would in fact be not physical at all but 
would simply coincide with the purely ``mathematical degrees of freedom'' of 
the {\it point constituents} of the arithmetical continuum used to formulate 
general relativity mathematically. We have to acknowledge that as long as the 
physical origin of black hole entropy is not confirmed by the experimental 
discovery of e.g. black hole radiation \cite{haw2,wal1} or other thermal 
phenomena, this possibility cannot be {\it a priori} refuted. The second 
extreme reading is that black hole entropy is of pure physical origin without 
any mathematical content which arises if one rather wishes to  
accept the physical origin of black hole entropy (and e.g. looks forward 
its experimental discovery). Then (\ref{vegeredmeny}) can be 
interpreted as an argument for the ``physical origin'' (cf. 
Heisenberg uncertainity) of what we have called the inherent fuzziness or 
uncertainty within the set of real 
numbers \cite{sch1,sch2}. This interpretation then could explain the 
expected independence of 
the set-theoretic entropy (\ref{entropia1}) of any axiomatic system $S$ which 
is in accordance with the universal character of Theorem \ref{chaitin}. 


\section{Appendix: a temporal approach to general relativity}
\label{four}


In this closing section we make an attempt to introduce temporality 
into general relativity by ``recycling'' the conceptional and technical 
apparatus introduced in Sections \ref{two} and \ref{three}. Quickly 
acknowledging that our efforts are certainly unsatisfactory yet, there is 
no doubt that we face a deep problem here taking into account the 
current paradoxical situation that although general relativity is the most 
advanced presently known physical theory about the structure of physical 
space and time, it is in fact a timeless theory due to its huge, namely 
diffeomorphism invariance (cf. \cite[Appendix E]{wal2}). For an excellent 
summary of the ``problem of time'' in general relativity see 
\cite[Chapter 3]{cal}. Hence a reconciliation is 
certainly necessary. 

However before sinking in technical details let us make some general remarks 
concerning the complex, diverse tension between the so-called manifest time and 
physical time. As we have recalled in Section \ref{one} and elaborated 
in Sections \ref{two} and \ref{three} in detail the division of the 
spatio-temporal continuum into disjoint points is conceptionally problematic. 
Encounters with quantum phenomena (in the form of {\it Heisenberg}'s 
uncertainty relations or various {\it EPR} phenomena, for instance) dictate 
that the physical space cannot be the simple union of extensionless disjoint 
empty ``places''; likewise and probably even more embarassingly 
psycho-physiological evidences (cf. e.g. \cite[Chapters 8 and 9]{cal} and 
the hundreds of related references therein) indicate that physical 
time cannot be the 
simple union of our ``nows''.  Indeed, nothing in the outer physical world 
seems to correspond to a (conscious) observer's inner now-experience. 
As {\it Carnap} recalls a personal conversation (\cite[pp. 37-38]{sch}): 

\begin{quotation}
\begin{small}

\noindent Once Einstein said that the problem of the Now worried him
seriously. He explained that the experience of the Now means something
special for man, something essentially different from the past and the
future, but that this important difference does not and cannot occur
within physics. That this experience cannot be grasped by science seemed
to him a matter of painful but inevitable resignation. I remarked that all
that occurs objectively can be described in science; on the one hand the
temporal sequence of events is described in physics; and, on the other
hand, the peculiarities of man's experiences with respect to time,
including his different attitude towards past, present, and future, can be
described and (in principle) explained in psychology. But Einstein thought
that these scientific descriptions cannot possibly satisfy our human
needs; that there is something essential about the Now which is just
outside of the realm of science.

\end{small}
\end{quotation}

\noindent What is so essential about the ``now''? One can effectively 
analyse the content of a conscious observer's now-experience for instance 
within the framework of {\it Husserl}'s phenomenology (see e.g. \cite{hus}). 
Phenomenologically (or rather very roughly) speaking it resembles a thick 
whirling cloud of strongly interacting elements whose primary representatives 
are the immediately given sensations of outer or inner events, various 
intentions, reflections and (spontaneously) recalled memories from the past 
constituting the ``now'' itself; however these primary contents have 
already been the sources for 
further primary sensations, the targets of further iterated intentions, 
reflections or have been simply shifting towards the past as memories, etc., 
etc., i.e. are already secondary contents as well carrying the characters of 
the past; in this way furnishing the now-experience with a sort of persistent 
standing-flowing feature. Or as {\it Weyl} (who actually studied 
philosophy under {\it Husserl} in his young ages) probably more clearly puts it 
\cite[p. 92]{wey}:

\begin{quotation}
\begin{small}

\noindent What I am conscious of is for me both a being-now and, in its 
essence, something which, with its temporal position, slips away. In this way 
there arises the persisting factual existent, something ever new which 
endures and changes in consciousness. What disappears can reappear; not, of 
course, as an experience which I have over again, but as content of an 
(accurate) memory, having become something past. In the objective picture 
which I form of the course of the life , such a past thing is to be opposed 
to what presently is as something earlier. So we can gather the following 
concerning objectively presented time:
\begin{itemize}

\item[1.] an individual point in it is non-independent, i.e., is pure 
nothingness when taken by itself, and exists only as a ``point of transition'' 
(which, of course, can in no way be understood mathematically);

\item[2.] it is due to this essence of time (and not to contingent 
imperfections in our medium) that a fixed time-point cannot be 
exhibited in any way, that always only an {\it approximate}, never 
an {\it exact} determination is possible.
\end{itemize}

\end{small}
\end{quotation}

\noindent Thus apparently at least in part the essential about the 
now as an experience is its standing-flowing, slipping-away or 
let us say its persisting incoming-outgoing character. This latter 
characterization might be more clearly formulated by saying that the ``now'' 
equally shares the properties of the future and the past; and if the future is 
fundamentally different in its nature from the past then the ``now'' is 
indeed something very difficult to grasp.

We proceed further along these lines because it is a basic experience 
that the future and the past are indeed very different: for example, one 
cannot remember the future but can remember the past. In an early paper 
\cite{wei} {\it von Weizs\"acker} suggests that this is because 
the {\it future} consists of physical {\it possibilities} but the {\it past} 
consists of physical {\it facts} (this is a refinement of the homogeneous 
timeless concept, the physical {\it event} of relativity theory) and calls this 
dichotomy as the {\it chronologicality of time}. Thus the appropriate 
mathematical theory to model the open future 
would be probability theory while to model the fixed past would be geometry. 
Therefore, accepting this, in order to grasp 
the phenomenon of the ``now'' mathematically one should seek a 
mathematical structure in which probability and geometry are in 
conjunction. It is very interesting that a structure of this kind exists 
and in fact is well-known: a {\it Riemannian manifold}. Indeed, a 
(compact oriented) Riemannian manifold both accommodates a geometric 
structure by its metric and a Kolmogorov probability space by the 
normalized positive volume-form of the same metric. It comes as a 
further surprise that exactly the same mathematical structure, the 
(pseudo-)Riemannian manifold plays a central role in formulating general 
relativity mathematically. Therefore it seems that without effort and in 
fact in a very conservative way we can reach our aim how to reconcile 
temporality and general relativity by revisiting the role of the 
Riemannian structure in this theory.

But if we want probability theory to enter the game too then 
the first question is how to interpret physically the probabilities here? 
The space-time continuum is modeled by a Lorentzian $4$-manifold $(N,h)$ in 
general relativity and its points are physically 
interpreted as physical events (without further specification at this 
level of generality). The relative positions between these points i.e. 
the physical events then can be examined with the tools of 
Riemannian geometry. In particular a Riemannian sub-$3$-manifold 
$(M,g)\subset (N,h)$ can be identified with the space in an ``instant'' and 
its points are the right-now-occuring physical events in general relativity 
which on the one hand admit usual characterization in terms of the 
Riemannian structure $(M,g)$ such as lengths, area, curvature, etc. 
describing their factual or ``standing'' properties. In addition consider the 
Kolmogorov probability space $(M,\ca_g,\frac{1}{V_g}\mu_g)$ induced by 
$(M,g)$ too and if $A\in\ca_g$ put $p_g(A):=\frac{1}{V_g}\int_A\mu_g$. If 
$x_0\in M$ is a distinguished physical event then its sharp localization 
is not possible; thus in accord with the 
{\bf Assumption--mathematical form} we put  
\vspace{0.1in}

\noindent{\bf Assumption--physical form.}
{\it The number $0\leqq p_g(A)\leqq 1$ is the probability that a 
distinguished right-now-occuring physical event $x_0\in M$ 
appears in the spatial region $A\subseteqq M$.}
\vspace{0.1in}

\noindent Thus, on the other hand, the same right-now-occuring physical events 
admit further characterization in terms of the probability structure 
$(M,\ca_g,\frac{1}{V_g}\mu_g)$ hence one can talk about 
their appearance probabilities in spatial regions, their expectation values, 
etc. too capturing their possibilitial or ``flowing'' properties as well. 

Let $(N,h)$ be a globally hyperbolic $4$ dimensional space-time; 
then recall (cf. e.g. \cite[Chapter 10]{wal2}) that there exists a 
spacelike sub-$3$-manifold $M\subset N$ called a {\it Cauchy surface} 
with induced Riemannian metric $g:=h\vert_M$ such that 
$(N,h)=D^-(M,g)\cup D^+(M,g)$ i.e. $(N,h)$ can be written as the union of 
the past and future domain of dependencies of $(M,g)$ and obviously 
$(M,g)=D^-(M,g)\cap D^+(M,g)$.    

\begin{definition} Let $(N,h)$ be a globally hyperbolic 
Lorentzian $4$-manifold or called simply a space-time and assume that it 
admits a Cauchy surface $M\subset N$ which is compact and oriented. 
(This implies that 
if $g:=h\vert_N$ denotes the induced Riemannian metric then $(M,g)$ is complete 
moreover if $V_g:=\int_M\mu_g$ is its volume then $0<V_g<+\infty$.) 
Consider the induced decomposition $(N,h)=D^-(M,g)\cup D^+(M,g)$. 

The triple 
\[\left(D^-(M,g),(M,g),\left(M,\ca_g,\frac{1}{V_g}\mu_g\right)\right)\] 
is a {\rm chronological space-time} where $D^-(M,g)\subset (N,h)$ is a 
Lorentzian $4$-manifold equal to the past domain of dependence of $(M,g)$ 
within $(N,h)$ and is called the {\rm past}, $(M,g)$ is a compact oriented 
Riemannian $3$-manifold and called the {\rm presence} and 
$\Big(M,\ca_g,\frac{1}{V_g}\mu_g\Big)$ is the Kolmogorov probability measure 
space induced by $(M,g)$ and called the {\rm future}.  

A chronological space-time $\left(D^-(M_i,g_i),(M_i,g_i),\left(M_i,
\ca_i,\frac{1}{V_i}\mu_i\right)\right)$ with $i=0$ {\rm precedes} the other 
one with $i=1$ if there exists an isometric embedding 
$D^-(M_0,g_0)\subseteqq D^-(M_1,g_1)$ of their pasts as 
Lorentzian $4$-manifolds and they are {\rm equivalent} if both precede 
each other i.e. if both $D^-(M_0,g_0)\subseteqq D^-(M_1,g_1)$ and $D^-(M_1,g_1)
\subseteqq D^-(M_0,g_0)$ holds hence $D^-(M_0,g_0)$ and $D^-(M_1,g_1)$ 
are isometric. 
\label{kronologikus}
\end{definition}
\vspace{0.1in}

\begin{remark}\rm 1. We conclude with a few comments on this definition.
Compared with the usual definition of a space-time as a whole and first 
of all {\it timeless} entity in general 
relativity, in a chronological space-time this timeless structure 
has been broken up from the point of view of an observer into a more familiar 
past-presence-future {\it temporal} structure. This is achieved 
simply by replacing the $\big(D^-(M,g), (M,g), D^+(M,g)\big)$ 
decomposition of $(N,h)$ with 
$\left(D^-(M,g),(M,g),\left(M,\ca_g,\frac{1}{V_g}\mu_g\right)\right)$. 
As a subtlety of the definition note that the present $(M,g)$ does not 
determine the past $D^-(M,g)$ because to obtain it one needs the {\it a priori} 
knowledge on the ambient space-time $(N,h)$. However 
at least in the case of a {\it compact oriented} Cauchy surface $M$ 
(a purely technical requirement) the present $(M,g)$ does uniquely detemine 
the future $\left(M,\ca_g,\frac{1}{V_g}\mu_g\right)$; this follows from a 
theorem of Moser \cite{mos} stating that the space $(M,\frac{1}{V_g}\mu_g)$ 
is independent of the particular $(M,g)$ providing the unit-volume 
density by its normalized volume-form $\frac{1}{V_g}\mu_g$. This directional 
(in)dependence is therefore the manifestation of a sort of temporal openness 
of a chronological space-time towards the future. 

2. Nevertheless, in spite of this future-openness, a globally 
hyperbolic space-time uniquely determines its associated chronological 
space-time. More precisely, the equivalence of two chronological space-times 
as defined here implies that their pasts $D^-(M_0,g_0)$ and $D^-(M_1,g_1)$ 
are isometric; moreover by its compactness the Cauchy surface $M_0$ is 
diffeomorphic to the other one $M_1$ hence even their futures are isomorphic 
as probability spaces again by Moser's theorem \cite{mos}. Consequently this 
definition meets the demand for introducing temporality into general 
relativity in a covariant i.e. diffeomorphism-invariant way, found to be 
non-fullfillable in general cf. e.g. \cite[Sections 2.3-2.6]{cal}. 

However to gain this covariance the technical assumption that the Cauchy 
surface $M$ be compact as in Definiton \ref{kronologikus} is essential. For 
example, there exists a space-time diffeomorphic to $\R^4$ admitting 
Cauchy surfaces which are even not homeomorphic \cite{new-cla}. One of 
them is the standard $\R^3$ while the other is a so-called Whithead space 
\cite{whi}: an open contractible $3$-manifold $W$ which is not homeomorphic 
(hence not diffeomorphic) to $\R^3$; 
there are uncountable many pairwise non-homeomorphic Whitehead spaces but it 
is known \cite{mcm} that every Whithead space $W$ satisfies that the product 
$W\times\R$ is always diffeomorphic to $\R^4$. Consequently, given this 
space-time, its two associated chronological space-times of this sort 
might have non-isomorphic futures. 

3. This framework permits one to exhibit a mathematical model for 
coming into existence. One can say that within a given chronological 
space-time $\left(D^-(M,g_0),
(M,g_0),\left(M,\ca_0,\frac{1}{V_0}\mu_0\right)\right)$ the {\it occurence} of 
a new geometry $(M,g_1)$ is a 
chronological space-time $\left(D^-(M,g_1),(M,g_1),\left(M,\ca_1,
\frac{1}{V_1}\mu_1\right)\right)$, if exists, such that 
$\left(D^-(M,g_0),(M,g_0),\left(M,\ca_0,\frac{1}{V_0}\mu_0\right)\right)$ 
precedes $\left(D^-(M,g_1),(M,g_1),\left(M,\ca_1,
\frac{1}{V_1}\mu_1\right)\right)$. 
The Kullback--Leibler entropy of $(M,g_1)$ relative to $(M,g_0)$ as 
defined in Theorem \ref{kl} then measures the lacking of 
knowledge on the geometry $(M,g_1)$ from the viewpoint of the present geometry 
$(M,g_0)$. This necessarily implies that $(M,g_1)$ is in the future of 
$(M,g_0)$ for any spacelike section in the past of $(M,g_0)$ always 
satisfies $(M,g_{-1})\subset D^-(M,g_0)$ hence is explicitly known already. 

\end{remark}

\vspace{0.1in}

\noindent{\bf Statements and Declarations}. The author is grateful to 
K.-G. Schlesinger and P. Vrana for the stimulating discussions many years ago. 
There are no conflicts of interest to declare 
that are relevant to the content of this article. The work meets all ethical 
standards applicable here. All the not-referenced results in this work are 
fully the author's own contribution. No funds, grants, or other financial 
supports were received. Data sharing is not applicable to this article as 
no datasets were generated or analysed during the underlying study.

\end{document}